\documentclass[
  aps,
  prl,
  reprint,
  preprintnumbers,
  showpacs,
  groupedaddress,
  amsmath,
  amssymb,
  floatfix]{revtex4-1}
\usepackage{graphicx,epsfig,dcolumn,multirow}
\usepackage{natbib}
\newcolumntype{d}[1]{D{.}{.}{#1}}

\RequirePackage{xspace}

\newcommand{\simgt}{\,\hbox{\lower0.6ex\hbox{$\sim$}\llap{\raise0.6ex\hbox{$>$}}}\,}
\newcommand{\simlt}{\,\hbox{\lower0.6ex\hbox{$\sim$}\llap{\raise0.6ex\hbox{$<$}}}\,} 

\newcommand{\ov}[1]{\overline{#1}}

\newcommand{\gev}{\ensuremath{\, \mathrm{GeV}}}

\newcommand{\epm}[2]{
 \raisebox{-0.5ex}{\shortstack[l]{$\scriptstyle+#1$\\$\scriptstyle-#2$}}}

\usepackage{pstricks}
\newcmykcolor{darkgreen}{1 0 0.6 0.5}  %% cyan magenta yellow black

\newcommand{\uli}{}
\newcommand{\martin}{}

\newcommand{\myFitter}{\textit{my}Fitter}
\newcommand{\Ocal}{\mathcal{O}}

\begin{document}
%% \linenumbers

\preprint{TTP12-034}

\title{
%======
Impact of a Higgs boson at a mass of 126 GeV on the standard model with three and four fermion generations
}

\author{Otto Eberhardt$^{\,a}$,
  Geoffrey Herbert$^{\,b}$, Heiko Lacker$^{\,b}$,\\
  Alexander Lenz$^{\,c}$,  Andreas Menzel$^{\,b}$, Ulrich 
  Nierste$^{\,a}$, and Martin Wiebusch$^{\,a}$  
\vspace{0.6cm}}

\affiliation{
\mbox{$^{a}$ Institut f\"ur Theoretische Teilchenphysik,
Karlsruhe Institute of Technology, D-76128 Karlsruhe, Germany,}
\mbox{email: otto.eberhardt@kit,edu, ulrich.nierste@kit.edu, 
              martin.wiebusch@kit.edu}\\                
\mbox{$^{b}$ Humboldt-Universit\"at zu Berlin,
                   Institut f\"ur Physik,
                   Newtonstr. 15,
                   D-12489 Berlin, Germany,\hspace{3em}}\\
\mbox{e-mail: 
	      geoffrey.herbert@physik.hu-berlin.de, lacker@physik.hu-berlin.de, 
	      amenzel@physik.hu-berlin.de}\\
\mbox{$^{c}$ CERN - Theory Divison,
              PH-TH, Case C01600, CH-1211 Geneva 23,
              Switzerland, and}\\
\mbox{~\hspace{3em} 
              IPPP, Department of Physics, University of Durham, DH1
              3LE, UK,  {e-mail: alenz@cern.ch}}
%% \mbox{$^{d}$ CKMfitter collaboration} 
}  
\date{October 22, 2012}

\begin{abstract}
  We perform a comprehensive statistical analysis of the standard model (SM)
  with three and four generations using the latest Higgs search results from LHC
  and Tevatron, the electroweak precision observables (EWPOs) measured at LEP
  and SLD and the latest determinations of $M_W$, $m_t$ and $\alpha_s$.  For the
  three-generation case we analyse the tensions in the electroweak fit by
  removing individual observables from the fit and comparing their predicted
  values with the measured ones. In particular, we discuss the impact of the
  Higgs search results on the deviations of the EWPOs from their best-fit
  values.  Our indirect prediction of the top mass is
  $\uli m_t=175.7\epm{3.0}{2.2}\gev$ at $68.3\%$ CL, in good agreement with the
  direct measurement.  We also plot the preferred area in the $M_W$-$m_t$ plane.
  The best-fit Higgs mass is $126.0\gev$.  For the case of the SM with a
  perturbative sequential fourth fermion generation (SM4) we discuss the
  deviations of the Higgs signal strengths from their best-fit values.  The
  $H\to\gamma\gamma$ signal strength now disagrees with its best-fit SM4 value
  at more than $4\sigma$. We perform a likelihood-ratio test to compare the SM
  and SM4 and show that the SM4 is excluded at $5.3\sigma $.  Without the
  Tevatron data on $H\to b \bar b$ the significance drops to $4.8\sigma$.
\end{abstract}

%========================================

\maketitle

%%%%%%%%%%%%%%%%%%%%%%%%%%%%%%%%%%%%%%%%%%%%%%%%%%%%%%%%%%%%%%%%%%%%%%%%%%%%%%%%
\section{Introduction}
%%%%%%%%%%%%%%%%%%%%%%%%%%%%%%%%%%%%%%%%%%%%%%%%%%%%%%%%%%%%%%%%%%%%%%%%%%%%%%%%

Electroweak precision observables (EWPOs) played an important role in the
prediction of the mass of the top quark prior to its discovery
\cite{Abe:1995hr,Abachi:1995iq}. Later, with improving Tevatron data on the mass
$m_t$ of the top quark, EWPOs were used to constrain the mass $m_H$ of the Higgs
boson, albeit with little precision since EWPOs only depend logarithmically on
this quantity \cite{Veltman:1976rt}. Recently ATLAS and CMS have discovered a
convincing candidate for the Higgs boson with a mass around 126 GeV
\cite{:2012gk,:2012gu}. With the information on $m_H$ available, EWPOs enter a
new era, as they directly test the standard model (SM) without involving
otherwise undetermined fundamental parameters. In this paper we present a
combined fit of the EWPOs and Higgs signal strengths in the decays to
$\gamma\gamma$, $WW$, $ZZ$, $b\bar b$, and $\tau\tau$ studied at the LHC and the
$p\ov p\to H \to b\bar b$ signal strength determined at the Tevatron
\cite{:2012zzl}.

The SM is minimal in several respects, e.g.\ the fermions belong to the
smallest possible representations of the gauge groups and electroweak
symmetry breaking is achieved with a single Higgs doublet. However, the
fermion field content is non-minimal and organised in at least three
families. There has been a tremendous interest in the phenomenology of a
fourth fermion generation, with more than 500 papers in the last decade.
The SM with a sequential fourth generation, SM4, had survived global
analyses of EWPOs and flavour observables
\cite{Frampton:1999xi,Kribs:2007nz,Holdom:2009rf,Chanowitz:2009mz,
  Erler:2010sk,Eberhardt:2010bm,Baak:2011ze}, but was put under serious
pressure from the first LHC data on Higgs searches
\cite{Djouadi:2012ae,Kuflik:2012ai,Eberhardt:2012sb,Buchkremer:2012yy,Eberhardt:2012ck}. In
this paper we show that the perturbative SM4 is the first popular model
of new physics which is ruled out by the LHC at the 5$\sigma$
level. This strong statement is possible because of the non-decoupling
property of the SM4, as loops with 4th-generation fermions do not vanish
with increasing masses.  For the same reason it is difficult to compute
the statistical significance at which the SM4 is ruled out: the
non-decoupling property implies that the SM4 and SM are non-nested,
i.e.\ the SM is not obtained from the SM4 by fixing the additional
parameters. This complicates the statistical procedure which compares
the performance of the two models at describing the data. We have solved
this problem with the help of a new method for toy Monte Carlo
simulations, as implemented in the {\myFitter} package
\cite{Wiebusch:2012en}. A first application of {\myFitter} to the SM4
with the {\martin data available before the announcement of the Higgs
  discovery {right before} the ICHEP 2012 conference} has been
  presented in \cite{Eberhardt:2012ck}.

%%%%%%%%%%%%%%%%%%%%%%%%%%%%%%%%%%%%%%%%%%%%%%%%%%%%%%%%%%%%%%%%%%%%%%%%%%%%%%%%
\section{Method and Inputs}
%%%%%%%%%%%%%%%%%%%%%%%%%%%%%%%%%%%%%%%%%%%%%%%%%%%%%%%%%%%%%%%%%%%%%%%%%%%%%%%%

We combine electroweak precision data with Higgs signal strengths provided by
ATLAS, CMS and Tevatron. {\martin (To study the impact of the excess in $H\to
  b\bar b$ events reported by CDF we also show results with the CDF input
  excluded.)} Our fit parameters in the SM are {\martin the $Z$ mass $M_Z$, the
  top quark mass $m_t$, the strong coupling $\alpha_s$, the hadronic
  contribution $\Delta\alpha_\text{had}^{(5)}$ to the fine structure constant at
  the scale $M_Z$ in the five-flavour-scheme and the Higgs mass $m_H$.} In the
SM4 the additional parameters are the fourth generation quark masses $m_{t'}$
and $m_{b'}$ and the fourth generation lepton masses $m_{\ell_4}$ and
$m_{\nu_4}$.  Mixing between the quarks of the fourth and the first three
generations is neglected in this analysis, since our previous study
\cite{Eberhardt:2012ck} showed that such mixing is disfavoured by the
combination of Higgs signal strengths and EWPOs. Direct searches for
fourth-generation quarks at the Tevatron and the LHC put lower bounds on the
masses of the fourth-generation quarks. For example, the current highest limit
on $m_{b'}$ is given in \cite{Chatrchyan:2012yea} as $m_{b'} > 611\gev$.
However, these limits rely on specific assumptions about the mass splitting and
the decay patterns of the heavy quarks.  We therefore use a conservative limit
of $m_{t'},m_{b'}>400\gev$ in our fits. For the lepton masses we require
$m_{\ell 4}>100\gev$ and $m_{\nu 4}>M_Z/2$.  As upper limit for all fourth
generation fermion masses we choose $800\gev$.

\begin{table}
  \centering
  \renewcommand{\arraystretch}{1.3}
  \begin{tabular}{lccc}
    \hline\hline
    Process & Reference(s) & Reference(s) & Combination \\
    & ($m_H$ free) & ($m_H$ fixed) & at 126\gev \\
    \hline
    $pp\to H\to\gamma\gamma$ & \cite{:2012gk}, \cite{CMS-PAS-HIG-12-015}
    & \cite{:2012gk}, \cite{CMS-PAS-HIG-12-015}& $1.583^{+0.337}_{-0.345}$\\
    $pp\to H\to WW^*$ & \cite{:2012gk}
    & \cite{:2012gk}, \cite{:2012gu} & $0.905^{+0.323}_{-0.294}$\\
    $pp\to H\to ZZ^*$ & \cite{:2012gk}
    & \cite{:2012gk}, \cite{:2012gu} & $0.861^{+0.391}_{-0.285}$\\
    $p\bar p\to HV\to Vb\bar b$ & \cite{:2012zzl}
    & \cite{:2012zzl} & $2.127^{+0.806}_{-0.763}$\\
    $pp\to HV\to Vb\bar b$ & -
    & \cite{:2012gk}, \cite{:2012gu} & $0.478^{+0.783}_{-0.680}$\\
    $pp\to H\to\tau\tau$ & \cite{Aad:2012an}
    & \cite{Aad:2012an}, \cite{:2012gu} & $0.100^{+0.714}_{-0.699}$\\
    \hline\hline
  \end{tabular}
  \caption{Experimental inputs for Higgs signal strengths.
  Except for $H\to\gamma\gamma$ CMS only provides signal strengths
  at $125.5\;$\gev. ATLAS has not published a 2012
  update on $H\to\tau\tau$ and $H\to b\bar b$, so we take the 2011 data.}
  \label{tab:signalstrengths}
\end{table}
Our inputs for the Higgs signal strengths are summarised in Table
\ref{tab:signalstrengths}. For a given Higgs production and decay mode $X\to
H\to Y$, the signal strength $\hat\mu(X\to H\to Y)$ is defined as the observed
production cross section times branching ratio divided by the SM prediction.
{\martin The asymmetric errors are accounted for by using an asymmetric gaussian
  likelihood function.}  For the SM fit and the Higgs mass scans we treat $m_H$
as a free parameter and interpolate the data from signal strength plots versus
Higgs mass, as provided by the ATLAS, CMS, CDF and D0 collaborations.  When
comparing the SM and the SM4 we keep the Higgs mass fixed in our fit and use the
combined signal strengths given in Table \ref{tab:signalstrengths} as inputs.

{\martin On the theory side, the SM4 signal strengths are computed by
  appropriately scaling the SM branching fractions and production cross sections
  separately for each production mechanism. (Further details can be found in
  \cite{Eberhardt:2012sb}.) In this sense, our treatment of the Higgs signal
  strengths is a special case of effective coupling analyses such as
  \cite{Lafaye:2009vr, Azatov:2012rd, Klute:2012pu, Carmi:2012in,
    Espinosa:2012im}.  Unfortunately, %% the published information about 
  {these
  analyses are} insufficient for constraining, let alone ruling out a 
  %% realistic
  {concrete}  
  model like the SM4.  An effective coupling analysis which ``contains'' the SM4
  would have to treat the Higgs couplings to $\gamma\gamma$, $WW$, $ZZ$, $gg$,
  $b\bar b$, $\tau\tau$ \emph{and} $\nu_4\bar\nu_4$ as independent parameters
  and provide full information about the $\chi^2$ function on this
  seven-dimensional parameter space. Even then %% it would be impossible to 
  {one could not} compute
  the $p$-value of the likelihood ratio test comparing the SM and the SM4, since
  this requires a numerical simulation with toy measurements. 
  % Hence, we do not
  % discuss the SM4 in terms of effective couplings but analyse its
  % compatibility
  % with the measured Higgs signal strengths and EWPOs directly.
  {Hence the results of \cite{Lafaye:2009vr, Azatov:2012rd,
      Klute:2012pu, Carmi:2012in, Espinosa:2012im} cannot be applied to
    the SM4.}}

Although our experimental inputs are combined results from the $7$ and
$8\;\text{TeV}$ LHC runs, we compute the signal strengths using $7\;\text{TeV}$
SM cross sections only. This practice is justified because the signal strengths
only depend on the ratios of Higgs production cross sections for different
production mechanisms and not on their absolute size. The ratios are constant to
a good approximation when going from $7$ to $8\;\text{TeV}$
{\martin\footnote{For example, the ratio of the gluon fusion and the VBF
    production cross sections is $12.577$ at $7\;\text{TeV}$ and $12.258$ at
    $8\;\text{TeV}$ for $m_H=126\;\text{GeV}$.}}.  Note however, that we treat
the $H\to b\bar b$ signal strengths from the Tevatron and LHC detectors as two
different observables because the ratio of $W$ and $Z$ associated production
cross sections is different at the Tevatron and at the LHC.

Heavy 4th generation fermions imply large Yukawa couplings which eventually make
the theory non-perturbative. The 1978 paper \cite{Chanowitz:1978mv} estimated a
breakdown of perturbation theory at $m_{b'} \geq 500 - 600\gev$ from
considerations of tree-level partial wave unitarity \cite{Lee:1977eg}. However,
this bound merely implies that for $m_{b'} \approx 500\gev$ loop corrections
become important. In our fits we compute the Higgs width and branching ratios in
the SM4 and SM with HDECAY v.\ 4.45 \cite{Djouadi:1997yw}, which implements the
higher-order corrections of \cite{Djouadi:1994gf, Djouadi:1994ge,
  Passarino:2011kv, Denner:2011vt} (see also \cite{Anastasiou:2010bt}).

The global fits with a variable Higgs mass were done with the CKMfitter software
\cite{Hocker:2001xe}. The EWPOs in the SM4 were calculated with the method
described in \cite{Gonzalez:2011he}, using FeynArts, FormCalc and LoopTools
\cite{Hahn:1998yk, Hahn:2000kx, Hahn:2006qw} to compute the SM4 corrections to
the EWPOs. The EWPOs in the SM were calculated with the ZFitter software
\cite{Bardin:1989tq, Bardin:1999yd, Arbuzov:2005ma}. The SM Higgs production
cross sections were taken from \cite{Dittmaier:2011ti} (LHC) and
\cite{Brein:2003wg, Baglio:2010um} (Tevatron). For the numerical computation of
the $p$-values we use the {\myFitter} package \cite{Wiebusch:2012en} which in
turn uses the Dvegas code \cite{Dvegas, Kauer:2001sp, Kauer:2002sn} for
numerical Monte Carlo integration.

%%%%%%%%%%%%%%%%%%%%%%%%%%%%%%%%%%%%%%%%%%%%%%%%%%%%%%%%%%%%%%%%%%%%%%%%%%%%%%%%
\section{SM fit results}
%%%%%%%%%%%%%%%%%%%%%%%%%%%%%%%%%%%%%%%%%%%%%%%%%%%%%%%%%%%%%%%%%%%%%%%%%%%%%%%%

\begin{table*}
  \centering
  \renewcommand{\arraystretch}{1.3}
  \begin{tabular}{llllll}
    \hline\hline
    Quantity & Input & Reference & Best fit value & Prediction & $\Delta \chi ^2$\\
    \hline
    $\sigma^0_\text{had}$[\ensuremath{\, \mathrm{nb}}] & $41.541\pm 0.037$ & \cite{PDG2012}
    & $41.4766^{+0.0075}_{-0.0141}$ & $41.468^{+0.014}_{-0.012}$ & $2.83$\\
    $A_\text{FB}^{0,l}$ & $0.0171\pm 0.0010$ & \cite{ALEPH:2005ab}
    & $0.016182^{+0.000073}_{-0.000079}$ & $0.016180^{+0.000072}_{-0.000081}$ & $0.90$\\
    $A_\text{FB}^{0,c}$ & $0.0707\pm 0.0035$ & \cite{ALEPH:2005ab}
    & $0.07357^{+0.00018}_{-0.00020}$ & $0.07357^{+0.00018}_{-0.00019}$ & $0.27$\\
    $A_\text{FB}^{0,b}$ & $0.0992\pm 0.0016$ & \cite{ALEPH:2005ab}
    & $0.10297^{+0.00023}_{-0.00025}$ & $0.10303^{+0.00023}_{-0.00024}$ & $4.74$\\
    $A_l$ & $0.1499\pm 0.0018$ & \cite{ALEPH:2005ab,Baak:2011ze}
    & $0.14689^{+0.00033}_{-0.00036}$ & $0.14679^{+0.00033}_{-0.00045}$ & $2.89$\\
    $A_c$ & $0.670\pm 0.027$ & \cite{ALEPH:2005ab}
    & $0.66781^{+0.00014}_{-0.00016}$ & $0.66781^{+0.00014}_{-0.00016}$ & $0.02$\\
    $A_b$ & $0.923\pm 0.020$ & \cite{ALEPH:2005ab}
    & $0.934643\pm 0.000025$ & $0.934643\pm 0.000025$ & $0.19$\\
    $R_l^0$ & $20.767\pm 0.025$ & \cite{ALEPH:2005ab}
    & $20.7420^{+0.0176}_{-0.0088}$ & $20.7365^{+0.0147}_{-0.0042}$ & $0.84$\\
    $R_c^0$ & $0.1721\pm 0.0030$ & \cite{ALEPH:2005ab}
    & $0.172249^{+0.000053}_{-0.000031}$ & $0.172249^{+0.000053}_{-0.000030}$ & $0.01$\\
    $R_b^0$ & $0.21629\pm 0.00066$ & \cite{ALEPH:2005ab}
    & $0.215804^{+0.000040}_{-0.000020}$ & $0.215803^{+0.000040}_{-0.000020}$ & $0.27$\\
    $\sin^2\theta_l^\text{eff}$ & $0.2324\pm 0.0012\pm 0.000047$ & \cite{ALEPH:2005ab,Baak:2011ze}
    & $0.231539^{+0.000045}_{-0.000041}$ & $0.231538^{+0.000044}_{-0.000042}$ & $0.46$\\
    $M_W$[\gev] & $80.385\pm 0.015\pm 0.004$ & \cite{TEW:2012gb,Baak:2011ze}
    & $80.3694^{+0.0049}_{-0.0072}$ & $80.3682^{+0.0051}_{-0.0135}$ & $0.66$\\
    $\Gamma_W$[\gev] & $2.085\pm 0.042$ & \cite{TEW:2010aj}
    & $2.09145^{+0.00113}_{-0.00086}$ & $2.09146^{+0.00113}_{-0.00087}$ & $0.02$\\
    $\Gamma_Z$[\gev] & $2.4952\pm 0.0023$ & \cite{ALEPH:2005ab}
    & $2.49561^{+0.00143}_{-0.00080}$ & $2.49532^{+0.00164}_{-0.00060}$ & $0.12$\\
    \hline
    $M_Z$[\gev] & $91.1876\pm 0.0021$ & \cite{PDG2012}
    & $91.1878^{+0.0020}_{-0.0021}$ & $91.192^{+0.014}_{-0.010}$ & $0.17$\\
    $m_t$[\gev] & $173.18\pm 0.56\pm 0.75$ & \cite{Aaltonen:2012ra}
    & $174.04^{+0.54}_{-1.14}$ & $175.7^{+3.0}_{-2.2}$ & $0.61$\\
    $\alpha_s (M_Z)|_\tau $ & $0.1202\pm 0.0006 \pm 0.0021$ & \cite{Baikov:2008jh}
    & $0.1189^{+0.0026}_{-0.0013}$ & $0.1189\pm 0.0027$ & $0.00$\\
    $\Delta\alpha_\text{had}^{(5)}(M_Z)$ & $0.02757\pm 0.00010$ & \cite{Davier:2010nc}
    & $0.027558\pm 0.000097$ & $0.02735^{+0.00042}_{-0.00047}$ & $0.26$\\
    $m_H$[\gev] & signal strengths & see Table \ref{tab:signalstrengths}
    & $126.00^{+0.36}_{-0.67}$ & $108^{+25}_{-33}$ & $6.26$\\
    \hline\hline
  \end{tabular}
  \caption{Experimental inputs and fit results for the electroweak precision
    observables in the SM. The observables were calculated with ZFitter
    \cite{Bardin:1989tq, Bardin:1999yd, Arbuzov:2005ma}. The inputs are listed
    in the second column.  The first error is statistical while the second (if
    present) is systematic.  We also use the correlations from
    \cite{ALEPH:2005ab}. The input for $\alpha_s$ is the value determined from
    the $\tau$ lifetime. In the fourth column, we show the results of a global
    fit using all available inputs. Here, the errors are $68.3\%$ CL
    intervals. The fifth column contains the prediction for each observable,
    obtained by removing the direct input for that observable and re-running the
    fit.  The corresponding difference of the minimal $\chi^2$ values is shown
    in column six.  The quantities in the last five rows were used as fit
    parameters.}
  \label{tab:ewpo}
\end{table*}
The EWPO inputs as well as the SM fit results can be found in Table
\ref{tab:ewpo}.  The experimental error on the Higgs mass is determined in a fit
to all available signal strengths (see the second column of Table
\ref{tab:signalstrengths}). In addition to the best-fit value and $68.3\%$ CL
interval of each observable we also show the prediction for each observable,
which is obtained by removing the direct input for that observable and
minimising the $\chi^2$ function again. The resulting change $\Delta\chi^2$ in
the minimum $\chi^2$ value is shown in the last column of Table
\ref{tab:ewpo}. A large $\Delta\chi^2$ value indicates that the observable is in
strong disagreement with the other observables. The most prominent ``outlier''
is therefore $A^b_\text{FB}$, followed by $\sigma^0_\text{had}$ and $A_l$. The
leptonic left-right asymmetry $A_l$ is the main cause for the disagreement
between the predicted and measured value of $m_H$: removing the $A_l$ input
leads to a predicted Higgs mass of $124\gev$. Note however, that the updated
inputs for $m_t$ and $M_W$ move the predicted Higgs mass up to $108\gev$.

\begin{figure}
  \includegraphics[width=0.46\textwidth,viewport=200 315 455 665,clip=true]%
                  {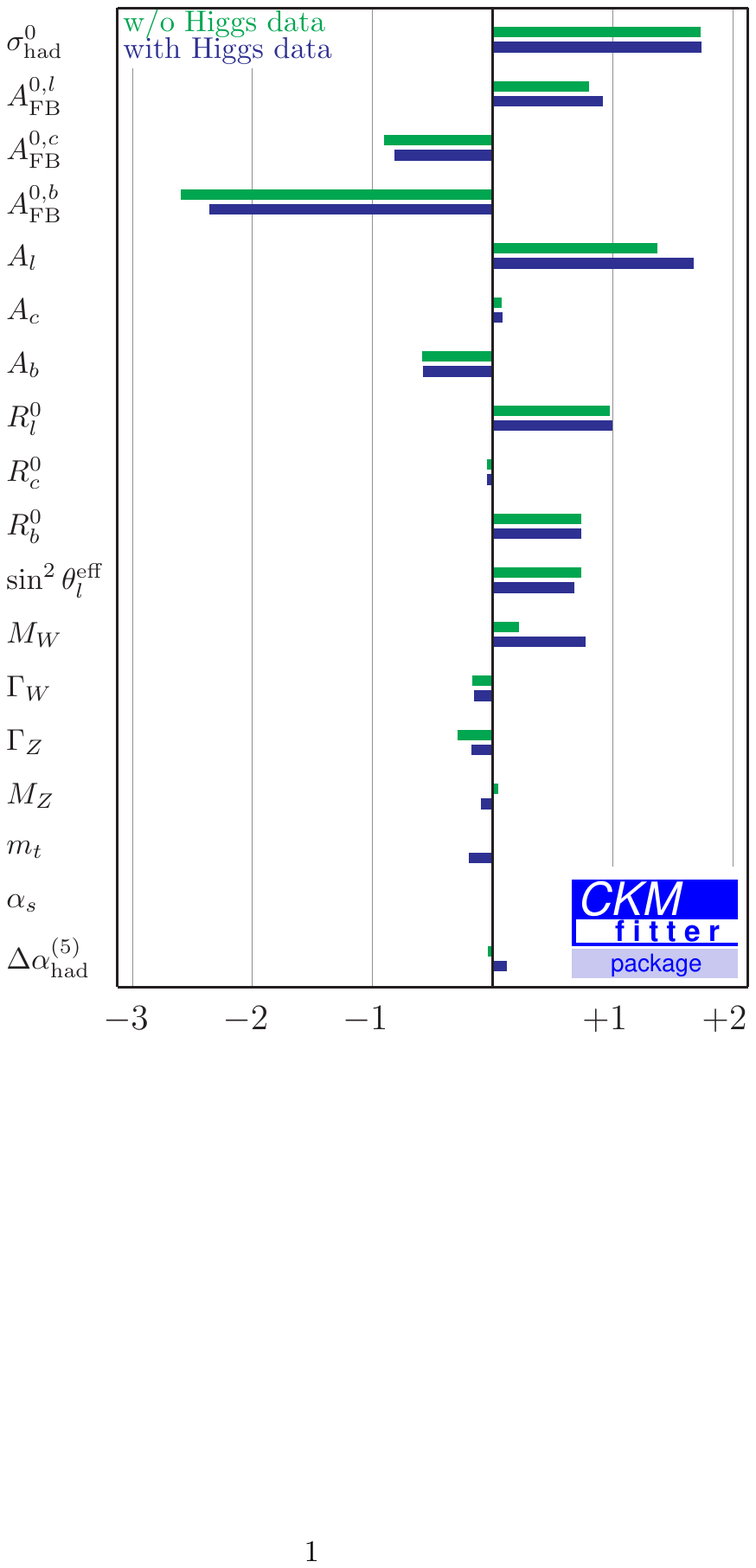}
  \caption{Deviations of the EWPOs in the Standard Model. The observables were
    calculated with ZFitter \cite{Bardin:1989tq, Bardin:1999yd,
      Arbuzov:2005ma}. For an observable $\Ocal$ with experimental value
    $\Ocal_\text{exp}$, experimental error $\Delta\Ocal_\text{exp}$ and best-fit
    prediction $\Ocal_\text{fit}$ we define the deviation as
    $(\Ocal_\text{exp}-\Ocal_\text{fit})/\Delta\Ocal_\text{exp}$.}
  \label{fig:SM3deviations}
\end{figure}
\begin{figure}
  \includegraphics[width=0.46\textwidth,viewport=192 419 474 604,clip=true]%
                  {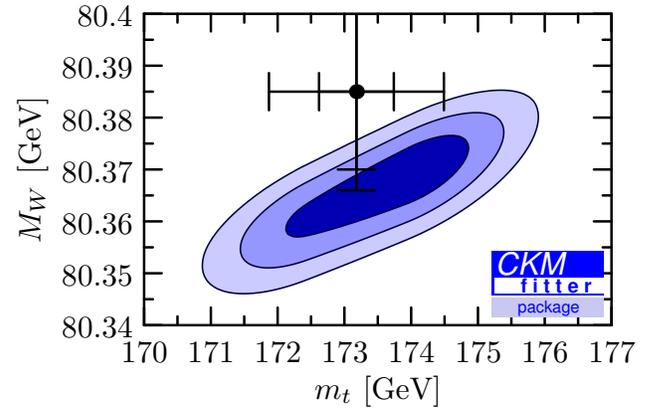}
  \caption{The $68.3\%$, $95.5\%$ and $99.7\%$ CL regions in the $m_t$-$M_W$
    plane using Higgs signal strengths and EWPOs. Also shown are the
    experimental values of $m_t$ and $M_W$ and their errors. The inner error
    bars are the statistical errors.}
  \label{fig:mtvsmW}
\end{figure}
The deviations of the EWPOs from their best-fit predictions are shown in
Figure~\ref{fig:SM3deviations}. This figure shows the fit with Higgs
signal strength and EWPO inputs as well as the fit with EWPO inputs
only.  Note that, due to the logarithmic dependence of the EWPOs on the
Higgs mass, the inclusion of the Higgs signal strengths is essentially
equivalent to fixing the Higgs mass at $126\gev$. We see that the new
Higgs data has a relatively small impact on the deviations of most
EWPOs.  The main difference is an increase in the deviation of $M_W$ to
$0.8\sigma$. The $68.3\%$, $95.5\%$ and $99.7\%$ CL regions in the
$m_t$-$M_W$ plane (using Higgs signal strengths and EWPOs) are shown in
Figure \ref{fig:mtvsmW}\footnote{{\uli Taking into account recently
  published higher-order corrections for $R_b^0$ \cite{Freitas:2012sy}
  changes the deviation of $R_b^0$ in Fig.~\ref{fig:SM3deviations} 
  to $+1.2$.  The effect on the
  best-fit parameters is marginal. For instance, the best-fit value of
  $m_t$ changes to $174.01\,\gev$.}}.

%%%%%%%%%%%%%%%%%%%%%%%%%%%%%%%%%%%%%%%%%%%%%%%%%%%%%%%%%%%%%%%%%%%%%%%%%%%%%%%%
\section{SM4 fit Results}
%%%%%%%%%%%%%%%%%%%%%%%%%%%%%%%%%%%%%%%%%%%%%%%%%%%%%%%%%%%%%%%%%%%%%%%%%%%%%%%%

The impact of a fourth fermion generation on the Higgs signal strengths has been
discussed extensively in the literature. The Higgs production cross section via
gluon fusion is enhanced by a factor of 9 due to the additional heavy quarks in
the loop \cite{Gunion:1995tp, Kribs:2007nz}. In $H\to\gamma\gamma$ searches,
this factor is overcompensated by a reduction of the branching ratios, which is
due to an accidental cancellation between gauge boson and fermion loops at
next-to-leading-order \cite{Denner:2011vt}. Finally, all signal strengths can be
suppressed by a common factor if the invisible $H\to\nu_4\bar\nu_4$ decay is
kinematically allowed \cite{Khoze:2001ug, Belotsky:2002ym, Bulanov:2003ka,
  Rozanov:2010xi, Keung:2011zc, Cetin:2011fp, Englert:2011us, Carpenter:2011wb}.

\begin{figure}
  \includegraphics[width=0.46\textwidth,viewport=141 455 425 708,clip=true]%
                  {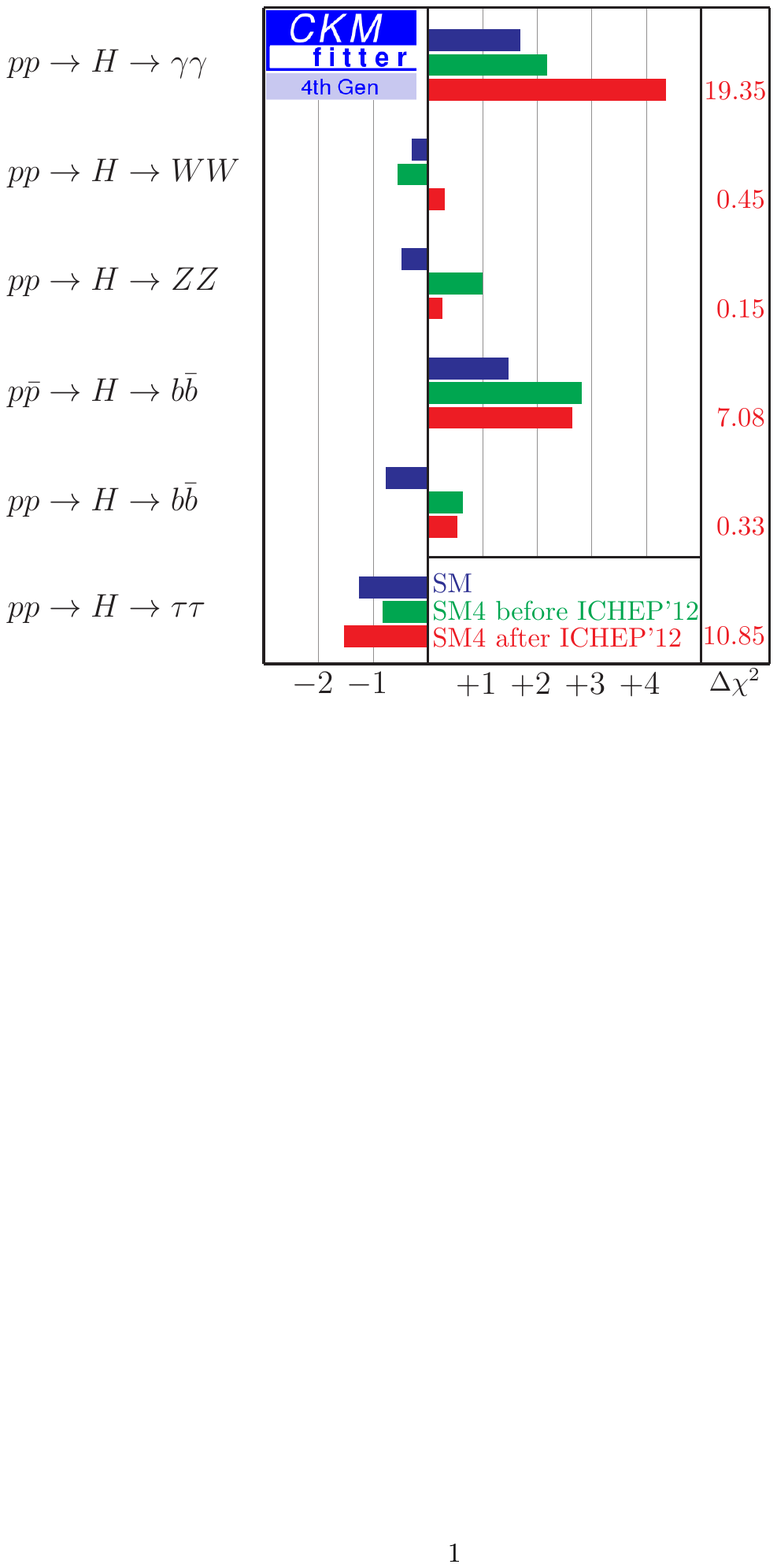}
  \caption{Deviations (defined as in Fig.~\ref{fig:SM3deviations}) of the Higgs
    signal strengths for the SM (blue) and for the SM4 (red) at a fixed Higgs
    mass of 126\gev. For comparison the results of the fit to pre-ICHEP2012 data
    from \cite{Eberhardt:2012ck} are also shown in green. In the right column we
    show, for the SM4 fit to current data, the change in the minimum $\chi^2$
    value when the corresponding signal strength is removed from the fit.}
  \label{fig:SM4pulls}
\end{figure}
The deviations of the Higgs signal strengths in the SM and the SM4 are shown in
Figure \ref{fig:SM4pulls}. For comparison we also show the deviations for the
SM4 fit to the data available before the announcement of the Higgs
  discovery, %% at the ICHEP 2012 conference
which was used in
\cite{Eberhardt:2012ck}.  For the SM4 fit to current data we also show the
shifts in the minimum $\chi^2$ value obtained by removing individual signal
strengths from the fit.  We see that the deviation of the $H\to\gamma\gamma$
signal strength has increased dramatically with the new data and now exceeds
four standard deviations.  Furthermore, the SM4 cannot explain an excess in
$H\to b\bar b$ searches because the Higgs production mechanisms for these
searches are $HW$ and $HZ$ associated production, which are not enhanced by a
factor of 9 like the gluon fusion production mode.  Thus the fit improves
significantly if the Tevatron measurement of the $H\to b\bar b$ signal strength
is removed.

\begin{figure}
  \includegraphics[width=0.46\textwidth,viewport=150 485 485 704,clip=true]%
                  {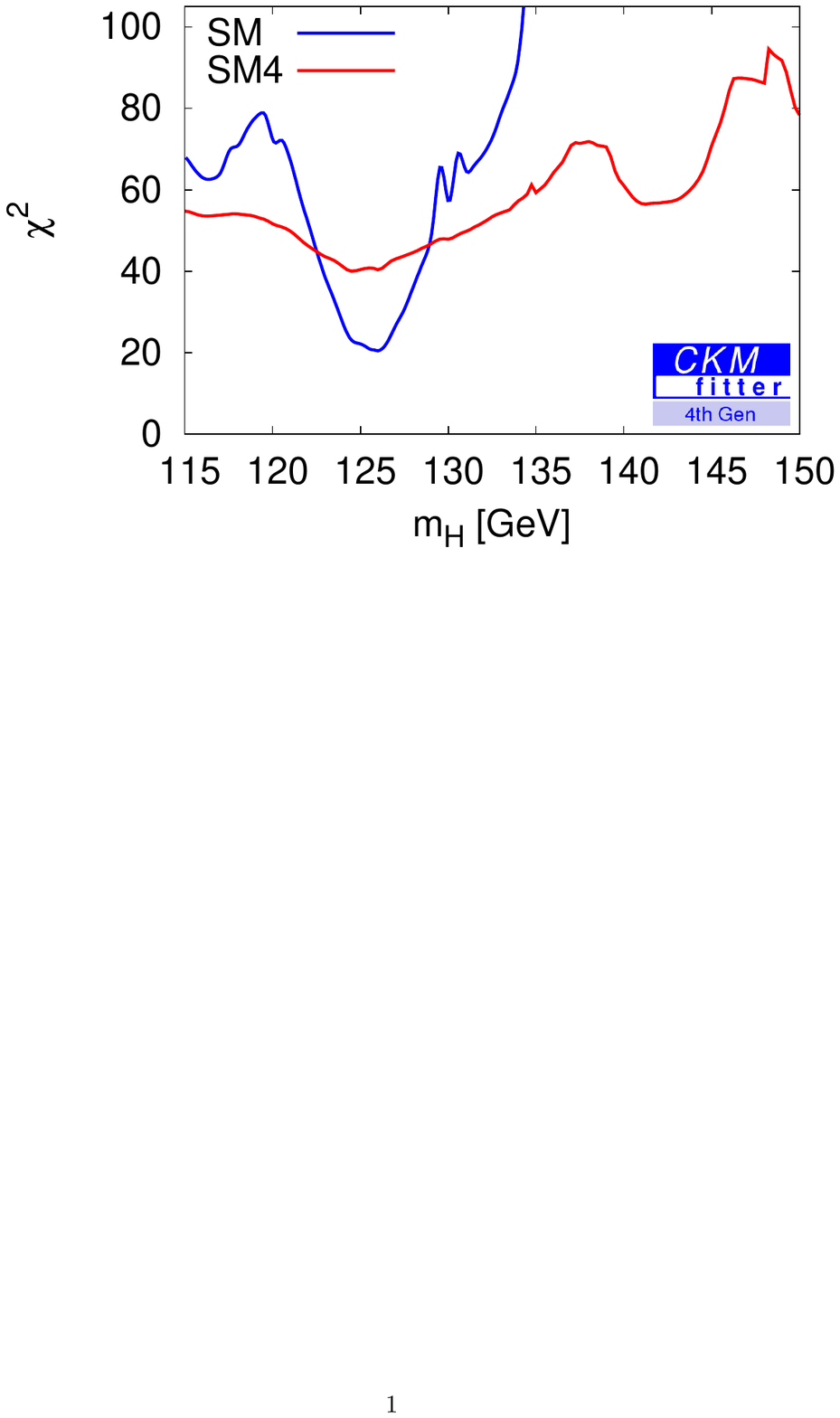}
  \caption{Higgs mass scan for the SM (blue line) and the SM4 (red line) based
    on the input set in the second column of Table \ref{tab:signalstrengths}.}
  \label{fig:mH}
\end{figure}
Figure \ref{fig:mH} shows the minimum $\chi^2$ values in the SM and the SM4 as
functions of the Higgs mass. The absolute minimum in the SM4 is at
$m_H=124.5\gev$ and the minimum $\chi^2$ value is larger than the one in the SM
by $20$ units.

To compute the statistical significance at which the SM4 is ruled out one has to
perform a likelihood-ratio test. This task is complicated by the fact that the
SM and the SM4 are not \emph{nested}, i.e.\ the extra parameters in the SM4
cannot be fixed in such a way that all observables assume their SM values.  As
explained in \cite{Wiebusch:2012en}, analytical formulae for $p$-values are not
valid in this case and one has to rely on numerical simulations. In our analysis
we used the improved simulation methods implemented in the {\myFitter} package.
For performance reasons, we fixed the SM parameters $M_Z$, $m_t$, $\alpha_s $,
$\Delta\alpha_\text{had}^{(5)}$ and $m_H$ to their best-fit values in these
simulations. This is a valid approximation since the SM4 fit is now dominated by
the Higgs signal strengths and their dependence on the SM parameters is
negligible. Table~\ref{tab:pvalues} summarises the results of the
likelihood-ratio tests. If all inputs are used, the SM4 is excluded at $5.3$
standard deviations. If the Tevatron input for the $H\to b\bar b$ signal
strength is removed the number of standard deviations drops to $4.8$.  Note that
these significances hold for an SM4 with a minimal Higgs sector and may be
weakened if the Higgs sector of the SM4 is extended \cite{BarShalom:2011zj,
  He:2011ti, BarShalom:2011bb, Chen:2012wz, Bellantoni:2012ag,
  BarShalom:2012he}.
\begin{table}
  \centering
  \renewcommand{\arraystretch}{1.3}
  \begin{tabular}{lcc}
    \hline\hline
    inputs                          & $p$-value         & standard dev.\\
    \hline
    all inputs                      & $1.1\cdot10^{-7}$ & 5.3 \\
    $H\to b\bar b$ from LHC only    & $1.9\cdot10^{-6}$ & 4.8 \\
    \hline\hline
  \end{tabular}
  \caption{The $p$-values and number of standard deviations of 
    likelihood-ratio tests comparing the SM and the SM4. The SM parameters
    were fixed to their best-fit values in these simulations.}
  \label{tab:pvalues}
\end{table}

%%%%%%%%%%%%%%%%%%%%%%%%%%%%%%%%%%%%%%%%%%%%%%%%%%%%%%%%%%%%%%%%%%%%%%%%%%%%%%%%
\section{Conclusions}
%%%%%%%%%%%%%%%%%%%%%%%%%%%%%%%%%%%%%%%%%%%%%%%%%%%%%%%%%%%%%%%%%%%%%%%%%%%%%%%%

We performed a combined fit of the parameters of the standard model with three
and four generations, combining Higgs search results and electroweak precision
data. In the SM electroweak fit the prediction for the Higgs mass from EWPOs has
moved closer to the value favoured by direct Higgs searches due to new inputs
for $m_t$ and $M_W$. When the Higgs signal strength inputs are combined with the
EWPOs the discrepancy between the measurement of $M_W$ and its best fit value
increases but stays below $1\sigma$. All other deviations are essentially
unaffected by the new input.

In the SM4 the measured $H\to\gamma\gamma$ signal strength disagrees with the
best-fit prediction by more than four standard deviations. Another source of
tension is the excess in $H\to b\bar b$ searches at the Tevatron in combination
with the deficit in $H\to\tau\tau$ events. The dominant Higgs production
mechanism for $H\to\tau\tau$ searches (gluon fusion) is enhanced by a factor of
$9$ in the SM4 while the relevant production mechanism for $H\to b\bar b$
searches ($HW$ and $HZ$ associated production) is slightly reduced.  The
statistical significance at which the SM4 is excluded must be computed by
numerical simulation methods like those implemented in the {\myFitter} software,
since analytical formulae for $p$-values do not hold in the case of non-nested
models. Using a conservative lower limit of $400\gev$ for the fourth-generation
quark masses and fixing the SM parameters to their best-fit values we find that
the SM4 with a minimal Higgs sector is ruled out at $5.3\sigma$.  If the results
of $H\to b\bar b$ searches at Tevatron are excluded from the analyses the SM4 is
still ruled out at $4.8\sigma$.

%%%%%%%%%%%%%%%%%%%%%%%%%%%%%%%%%%%%%%%%%%%%%%%%%%%%%%%%%%%%%%%%%%%%%%%%%%%%%%%%
\section{Acknowledgements}
%%%%%%%%%%%%%%%%%%%%%%%%%%%%%%%%%%%%%%%%%%%%%%%%%%%%%%%%%%%%%%%%%%%%%%%%%%%%%%%%

We thank Gr{\'e}gory Schott, Johann K\"uhn and J\'er\^ome Charles for fruitful
discussions.  We acknowledge support by DFG through grants NI1105/2-1,
LA2541/1-1, LE1246/9-1, and LE1246/10-1.

\bibliography{sm4higgs3}

\end{document}